\documentclass[a4paper,11pt]{article}
\pdfoutput=1 % if your are submitting a pdflatex (i.e. if you have
\usepackage{jcappub} % for details on the use of the package, please
\usepackage[english]{babel}
\usepackage{aas_macros} % journal abbreviations from ADS

\usepackage[T1]{fontenc} % if needed
\usepackage{subcaption}
\usepackage{amsmath, amssymb}
\usepackage{mathrsfs}
\usepackage{xspace}
\usepackage{blindtext}
\usepackage{textcomp}
\usepackage{algorithm}
\usepackage[export]{adjustbox}% http://ctan.org/pkg/adjustbox
\usepackage[noend]{algpseudocode}

\newcommand{\dd}{\text{d} }
\newcommand{\rmsub}[1]{_{\mathrm{#1}}}

\title{\boldmath A Probabilistic Model for the Efficiency of Cosmic-Ray Radio Arrays}

\author[a]{Vladimir~Lenok}
\author[a,b]{and Frank\,G.~Schr\"oder}

\affiliation[a]{Institute for Astroparticle Physics, Karlsruhe Institute of Technology (KIT),\\ 76021 Karlsruhe, Germany }
\affiliation[b]{Bartol Research Institute, Department of Physics and Astronomy, University of
Delaware,\\Newark DE, 19716, USA}

\emailAdd{vladimir.lenok@kit.edu}
\emailAdd{frank.schroeder@kit.edu}

\abstract{Digital radio detection of cosmic-ray air showers has emerged as an alternative technique in high-energy astroparticle physics.
Estimation of the detection efficiency of cosmic-ray radio arrays is one of the few remaining challenges regarding this technique. 
To address this problem, we developed a model based on the explicit probabilistic treatment of key elements of the radio technique for air showers: the footprint of the radio signal on ground, the detection of the signal in an individual antenna, and the detection criterion on the level of the entire array.
The model allows for estimation of sky regions of full efficiency and can be used to compute the aperture of the array, which is essential to measure the absolute flux of cosmic rays. 
We also present a semi-analytical method that we apply to the generic model, to calculate the efficiency and aperture with high accuracy and reasonable calculation time.
The model in this paper is applied to the Tunka-Rex array as example instrument and validated against Monte Carlo simulations. 
The validation shows that the model performs well, in particular, in the prediction of regions with full efficiency. 
It can thus be applied to other antenna arrays to facilitate the measurement of absolute cosmic-ray fluxes and to minimize a selection bias in cosmic-ray studies.}

\begin{document}
\maketitle
\flushbottom

\section{Introduction}

Observations of cosmic rays via the radio emission of air showers they initiate is one of the promising techniques for the next generation of ultra-high-energy astroparticle detectors.
The technique of digital radio detection has been under intensive development for the last twenty years and has reached the state where we can reliably detect radio signals from air showers and reconstruct their parameters~\cite{2016PhR...620....1H, 2017PrPNP..93....1S, 2022arXiv220505845C}.
One of the major remaining problems for radio arrays consists in the estimation of their detection efficiency for air showers and the subsequent calculation of their aperture and exposure for cosmic rays.

The method of Monte Carlo simulations, which is often used for these purposes, is difficult to apply for radio arrays because of its high computational complexity in case of the air-shower radio emission.
Compared to the approximately axially symmetric lateral distributions of the air-shower particles or the air-Cherenkov light emitted by air showers, the radio signal is more complex. 
Due to the interplay of geomagnetic and charge-excess emission, the strengths of the radio signal and the two-dimensional shape of the lateral distribution depend on both, the azimuth and zenith angle.
Also, Monte Carlo simulations of the radio emission of air showers take an order of magnitude more computation time than simulating the particles alone. 

Past approaches have either used a preliminary version of the model presented here~\cite{2018PhRvD..97l2004B, 2019ICRC...36..319K}, or estimated the efficiency for each detected event separately by generating several computationally expensive Monte Carlo simulations per event \cite{2015ICRC...34..368B, 2017ICRC...35..499B, 2021PhRvD.103j2006C}.
The preparation of simulations required for detailed studies of the spatial and angular dependence of the detection efficiency would demand an extremely large amount of computation time.

To address the problem of detection efficiency of a radio array, we developed a model following an explicit probabilistic approach.
For each step of the air-shower detection process (prediction of the spatial distribution of the radio signals over the array, detection of the signals by individual antennas, and the detection of the shower on the level of the array) we have developed a probabilistic model expressed as a combination of dedicated terms of a probability density function. 
The combination of these functions forms the final efficiency model.

In this work we used the Tunka-Rex digital radio antenna array as an example array for the model.
Despite some features of this particular array used here, the developed approach is generic and, with appropriate modifications of the model components, can be applied to other cosmic-ray radio arrays.
The paper is organized as follows: we first describe Tunka-Rex and the details of the simulation dataset used to build the model presented here; then we describe the efficiency model and its individual components; finally, we show how the aperture of a radio array can be computed semi-analytically.
In the end of the paper, we show the results of the validation of the model against Monte Carlo simulations.

\section{Tunka-Rex array}
The model we present here is built for the Tunka Radio extension (Tunka-Rex) as example radio array.
Tunka-Rex was a digital radio array for cosmic-ray detection in the Tunka valley in Siberia at the altitude of 675\,m above sea level (corresponds to 955\,g/cm$^2$ vertical atmospheric depth)~\cite{2015NIMPA.802...89B}.
The array was built in three stages and in its final configuration consisted of 63 antennas of SALLA type~\cite{2012JInst...7P0011A} covering approximately $3\,$km$^2$, with an inner dense core of almost $1\,$km$^2$.
Upon a trigger from the co-located arrays Tunka-133 and \mbox{Tunka-Grande}~\cite{2019JPhCS1263a2006K}, all Tunka-Rex antennas recorded a radio trace in the band of $30$--$80\,$MHz.
For cosmic rays with energies $\gtrsim 100\,$PeV, depending on the position of the antenna relative to the shower axis and on the arrival direction, the radio signal of the air showers can be distinguished from the Galactic noise and other local radio background.
Due to the continuous and omnipresent radio background, and due to the signal strengths of the radio emission depending on many parameters of the air showers and the relative position of the antennas, the detection process has a probabilistic nature that is captured by the model presented in this paper.
Details of the Tunka-Rex instrument, its operation, and the standard analysis pipeline of radio signals emitted by air showers can be found in Refs.~\cite{2015NIMPA.802...89B, 2016JCAP...01..052B, 2018PhRvD..97l2004B}.

\section{Simulation Dataset}
The efficiency model in this paper is based on full-fledged, end-to-end Monte Carlo simulations of the cosmic-ray air-shower radio emission, including the detector response of the Tunka-Rex instrument.
The simulations of air showers were prepared with CORSIKA~v7.5600 and v7.6400~\cite{Heck:1998vt} (there is no relevant differences regarding radio signal between these versions) using the QGSJET-II-04~\cite{2011PhRvD..83a4018O} and FLUKA~\cite{fluka} models for the high- and low-energy hadronic interactions correspondingly, and with the NKG and EGS4~\cite{Nelson1985, Heck:1998vt} models for the electromagnetic interactions.
The simulated showers are initiated by protons and iron nuclei as primary particles.
A discrete set of primary energies was chosen to enable sufficient statistics per energy bin: $\lg(E/\mathrm{eV})=17.0,~17.3,~17.5,~17.7,~18.0,~18.3$.
The simulation library prepared for this study consisted of approximately 1000 events per each energy bin and per each of the primary particles with uniform coverage of the incoming directions up to a zenith angle of 50$^{\circ}$ and uniform distribution of the shower cores.
The geomagnetic field for the simulations was set corresponding to the Tunka-Rex location with horizontal and vertical components of $\approx$18.88\,\textmu T and $\approx$57.29\,\textmu T, respectively, and the angle of declination of $-2.76^{\circ}$~\cite{WMM2015}.
The core position was varied randomly over the array with the actual antenna positions.
The radio emission coming from the simulated shower was computed with CoREAS~\cite{2013AIPC.1535..128H}.
Finally, the 
\mbox{$\overline{\textrm{Off}}$\hspace{.05em}\protect\raisebox{.4ex}{$\protect\underline{\textrm{line}}$}}
software~\cite{2007NIMPA.580.1485A} was used, first, to apply the detector response of the Tunka-Rex instrument and add on-site measured background and, second, to check with the standard analysis pipeline in which antennas the radio signal would be detected.

\section{Efficiency Model for a Radio Array}
\label{sec:model}

The efficiency model is based on an explicit probabilistic treatment of the spatial signal distribution, detection of these signals by individual antennas, and the detection condition on the level of the entire array (e.g., a certain number of antennas with a minimum signal-to-noise ratio).
By explicit probabilistic treatment we mean that we describe each of these stages of the detection process by a specific probabilistic model. 
Their combination forms the efficiency model for the radio array.

In the following sections we present the individual components of the model and the ways they interplay with each other.

\subsection{Spatial Distribution of the Air-Shower Radio Signals}

In our model, the estimation of the detection efficiency of a radio array begins with the evaluation of the spatial distribution of the signal strength corresponding to a shower with a
given set of macroscopic parameters, i.e., incoming direction, energy of the primary particle, and depth of the shower maximum ($X_{\text{max}}$).
This distribution is described by a lateral distribution function (LDF). 
Usually, such functions are used to estimate the shower parameters from the observed signal distribution. 
We use them in the opposite way: to predict the spatial distribution of the radio emission by showers with given macro-parameters.

Because of the geomagnetic and charge-excess emission mechanisms both being relevant for the frequency range from $30$ to $80\,$MHz used at Tunka-Rex and other arrays~\cite{1971NPhS..233..109P, 2014JCAP...10..014S, 2014PhRvD..89e2002A, 2016PhRvD..94j3010S, 2017PrPNP..93....1S}, the spatial distribution of the radio emission is axially-asymmetric relative to the shower axis.
The main idea behind the Tunka-Rex analysis, which guided the design of the
corresponding LDF, consists in the compensation of the asymmetry to perform the reconstruction with a one dimensional symmetrized LDF.
Thus, the Tunka-Rex LDF consists of two major components: the asymmetry-compensation operator and the symmetrized LDF, which we call LDF hereafter for simplification \cite{2016APh....74...79K}.

The symmetrized LDF, which we use in the reconstruction, has a Gaussian form expressed in the following way
\begin{equation}
  \label{eq:ldf-generic}
  \mathscr{E}\rmsub{sym}(r) \propto
  \exp\left( a\left(r-r'\right)^2 + b\left(r-r'\right) \right).
\end{equation}
The symbol $r'$ denotes a reference distance for which the parameters $a$ and $b$, and the normalization of that function are defined. 
The Tunka-Rex reconstruction uses two reference distances $r'$, which we denote as $r_0$ and $r_1$ ($r_0=120\,$m and $r_1=180\,$m). 
The value of the function at $120\,$m is approximately proportional to the cosmic-ray energy ($E$). 
The slope defined as a logarithmic derivative of the function at $180\,$m is related to the depth of shower maximum. 

We begin the construction of the footprint model with the LDF in the form (\ref{eq:ldf-generic}) with the
reference distance used for $X_{\text{max}}$ estimation ($r_1 = 180$\,m)
\begin{equation}
  \mathscr{E}_{\text{sym}}(r) \propto
  \exp\left( a\left(r-r_1\right)^2 + b\left(r-r_1\right) \right).
\end{equation}
For the parameters $a$ and $b$ we use the same equation as used in the reconstruction, but rearrange them to express the parameters
\begin{align}
  a &= \left(a_0 + a_1 E\right)
  + \left(a_2 + a_3 E\right)\cos\theta, \\
  b &= b_0 -
  \exp\left(
    \frac{1}{b_1}\left(
    \frac{X\rmsub{det}}
         {\cos\theta} - X_{\text{max}} - b_2\right)\right).
\end{align}
Here, the letters $a_i$ and $b_i$ denote the parameters obtained from a simulation study. 
The vertical atmospheric depth of the radio instrument is denoted as $X\rmsub{det}$ (955\,g/cm$^2$ for the present model).

The last step in deriving the LDF is to find an appropriate normalization. 
The value of the LDF at $r_0 = 120$\,m is proportional to the energy of the cosmic-ray with a calibration coefficient
$\mathscr{E}_{\text{sym}}(r_0) = E/\kappa$. 
To link this energy calibration at $r_0$ to the LDF defined at the distance $r_1$, we introduce an additional exponential
factor. 
The final, normalized LDF has the following form:
\begin{equation}
  \mathscr{E}_{\text{sym}}(r) =
  \frac{E}{\kappa}
  \exp\left( -a\left(r_0-r_1\right)^2 - b\left(r_0-r_1\right) \right)
  \exp\left( a\left(r-r_1\right)^2 + b\left(r-r_1\right) \right).
  \label{eq:ldf}
\end{equation}

To restore the original asymmetry of the radio footprint, we  act on the symmetrized LDF shown above with an inverse version of the operator used for the asymmetry compensation in the reconstruction
\begin{equation}
  \hat{\mathsf{K}}^{-1}(\alpha\rmsub{g}, \phi\rmsub{g})
  = \sqrt{c_0^2
    +2c_0\cos\phi\rmsub{g}\sin\alpha\rmsub{g}
    +\sin^2\alpha\rmsub{g}}.
\end{equation}
The letters $\alpha\rmsub{g}$ and $\phi\rmsub{g}$ denote the geomagnetic angle and the geomagnetic azimuth.
The first one is the angle between the shower axis and the geomagnetic field.
The later one is the polar angle in the shower plane of the geomagnetic coordinate system measured from the $\boldsymbol{V}\times\boldsymbol{B}$ direction.
With this operator acting on the symmetrized LDF (\ref{eq:ldf}), we obtain the original asymmetric form of the radio
footprint
\begin{equation}
  \mathscr{E}(r, \alpha\rmsub{g}, \phi\rmsub{g})
  = \hat{\mathsf{K}}^{-1}(\alpha\rmsub{g}, \phi\rmsub{g})
  \mathscr{E}\rmsub{sym}(r).
\end{equation}

\begin{figure}[!t]
  \includegraphics{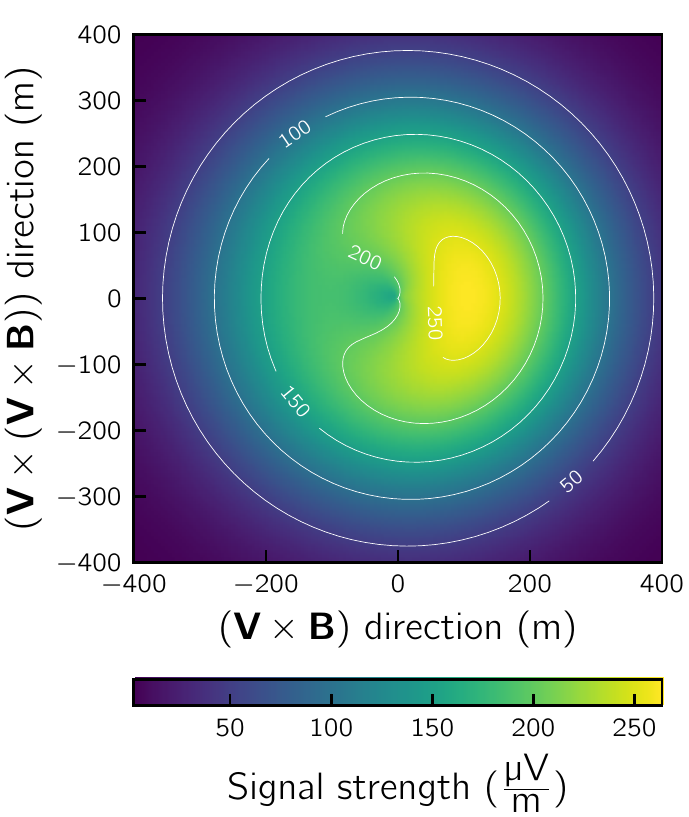}
  \hspace{0.05\textwidth}%
  \begin{minipage}[b]{0.455\textwidth}
    \caption{Spatial distribution of the most probable value (mode) of the
      electric field at $30$ to $80\,$MHz from an air shower with the following
      parameters:
      \mbox{$\lg(E/\text{eV})=17.5$},
      $X_{\text{max}}=400\,$g/cm$^2$,
      \mbox{$\theta = 40^{\circ}$},
      \mbox{$\alpha_g = 30^{\circ}$}.
      The shown distribution is the Tunka-Rex asymmetric lateral distribution. 
      The plot shows the distribution in the geomagnetic coordinate system, which is built from the vectors of the shower propagation direction $\boldsymbol{V}$ and the local geomagnetic field direction $\boldsymbol{B}$. 
      To obtain the distribution on the antenna array, the shown distribution is projected geometrically to the ground plane.
    }%
    \label{fig:footprint}
  \end{minipage}
\end{figure}

Figure~\ref{fig:footprint} shows a particular example of the distribution obtained with the procedure described above.

It is important to note that within the developed model the radio signal from a given shower does not have one specific signal strength at a given position, but instead the strength is a random variable. 
The reason for this random behavior is the fact that our description only includes effects related to the cosmic-ray energy, the depth of shower maximum, and the arrival direction.
However, the signals are also subject to natural shower-to-shower fluctuations leading to the randomization.

To find parameters of the probability distribution of the signal strength, we compared the prediction against the CoREAS simulations of individual showers without adding noise. 
Since the asymmetry correction makes only a linear transformation of the footprint, we used the symmetrized footprints for this purpose of determining the effect of the shower-to-shower fluctuations.  
A statistical analysis of the differences revealed that the developed footprint model provides the most probable value (mode) of the distribution and that the width of the distribution can be characterized with a standard deviation equal to $\approx14.5\,\%$ of the current mode of the distribution ($\sigma \approx 0.145\mathscr{E}$).
This spread originates from the shower-to-shower fluctuations since we used noiseless signals at this stage.
% influence of noise is considered at the next stage of the individual antenna detection probability.
%fluctuations due to noise are considered at a later stage.
We model this distribution with a Gaussian function centered at the mode value predicted by the footprint model and with the standard deviation determined by the statistical analysis mentioned above.
The additional influence of noise to the detection procedure is part of the next stage of modelling the individual antenna detection probability.

\subsection{Signal Detection by a Single Antenna}

The main measurement devices of a cosmic-ray radio array are the antennas which detect electric fields and convert them into currents 
which, in turn, can be detected by corresponding electronic devices.
In addition to the signals from the air showers, the antennas are always subject to the continuous, unavoidable presence of background electric fields, or simply noise.
In the band of $30$ to $80\,$MHz, this noise originates mainly from the radio sources in our Galaxy~\cite{ITUNoise2019} and the surroundings of the antenna. 
Noise interferes with the signals from the shower and, due to its stochastic nature, randomly changes the signal characteristics detected by the antenna.
Close to the detection threshold, this effect is the main reason for the probabilistic behavior of the signal detection. 
Due to the interference with noise, in some cases the presence of a signal is not detected by the system, or vice-verse a signal below threshold may be detectable due to an upward fluctuation.
We formulate these effects in terms of a probability density.

We start constructing the probability density of the signal detection by processing simulated radio signals multiple times through the Tunka-Rex signal processing pipeline. 
Each time a different noise sample is added to the simulation.
Noise samples used in this procedure were recorded by the Tunka-Rex array.
For each individual CoREAS simulation, we obtain the number of times a given signal was detected from the total number of trials (30 for our study), where each trial corresponds to a different measured noise sample. 
We estimate the detection probability for a given signal as the binomial proportion of these two numbers.

As next step, to obtain the continuous values of the detection probability as a function of the signal strength from the discrete values obtained previously, we fit the logistic function in the form of the hyperbolic function with an offset $\mathscr{E}_{1/2}$ to the obtained discrete values
\begin{equation}
  p_0(\mathscr{E}) = \frac{1}{2} +
  \frac{1}{2}\tanh\frac{\mathscr{E} - \mathscr{E}_{1/2}}{\mathscr{E}'_0 + \mathscr{E}''_0 \mathscr{E}}.
  \label{eq:mode}
\end{equation}
To provide sufficient degrees of freedom to match the data, we introduced a linear function to the denominator of the tangent argument. 

Now we will treat this detection probability not as a number, but as a random variable.
We model the probability density with the beta distribution in which the quantity $p_0$ found above (Equation (\ref{eq:mode})) describes the mode
\begin{equation}
  P = \frac{1}{\mathrm{B}(\alpha(p_0), \beta(p_0))}\,
  p^{\alpha(p_0)-1}(1-p)^{\beta(p_0)-1}.
  \label{eq:probability-density}
\end{equation}
The letter $\mathrm{B}$ denotes the beta function.
The beta function in this case can be seen as a continuous analogue of the binomial distribution.
The parameters of the distribution are linked to the mode $p_0$ of the distribution and the total number of trials $n$
\begin{align}
  \label{eq:pars1}
  \alpha(\mathscr{E}) &= n p_0(\mathscr{E}) + 1, \\
  \label{eq:pars2}
  \beta(\mathscr{E})  &= n - n p_0(\mathscr{E}) +1.
\end{align}

We obtain the parameters of the probability density~(\ref{eq:mode}--\ref{eq:pars2}) with a regular optimization procedure
based on the logarithmic-likelihood function, which we form from the beta distribution described above
\begin{equation}
  \mathcal{L} = \sum_i
  (\alpha_i - 1) \ln p_0(\mathscr{E}_i) +
  (\beta_i - 1)  \ln (1-p_0(\mathscr{E}_i)) -
  \ln \mathrm{B}(\alpha_i, \beta_i).
\end{equation}
The index $i$ refers to a single simulation data point. 
The symbol $\mathscr{E}_i$ denotes the signal strength of a given data point.
The parameters $\alpha_i$ and $\beta_i$ corresponding to each of the data points we determine as
\begin{align}
  \alpha_i &= k_i + 1, \\
  \beta_i  &= n -k_i +1,
\end{align}
where $k_i$ 
denote the number of successful signal detections.

\begin{figure}[p]
  \centering
  \includegraphics{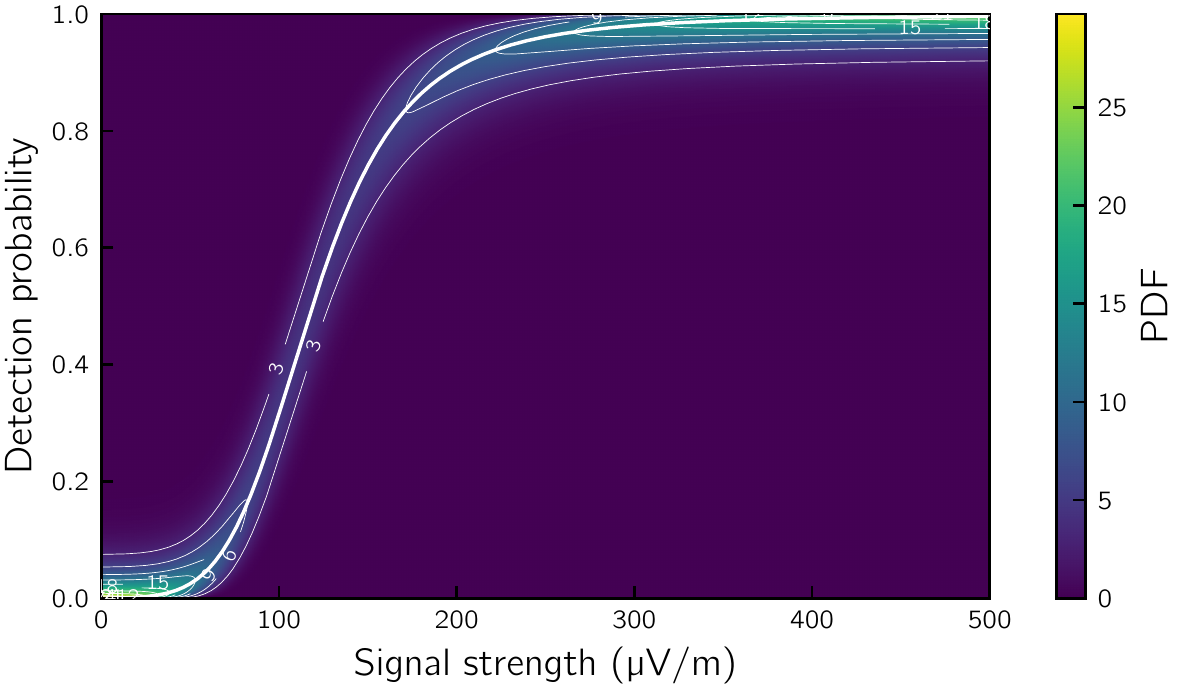}
  \caption{The probability density function (PDF) of the signal detection by an individual antenna (corresponds to Equation~\ref{eq:probability-density}). 
  The thick line shows the mode value.}
  \label{fig:pdf}
\end{figure}

Figure~\ref{fig:pdf} shows the estimated density of the detection probability for a signal with known strength by an
individual antenna.

\begin{figure}[p]
  \centering
  \includegraphics{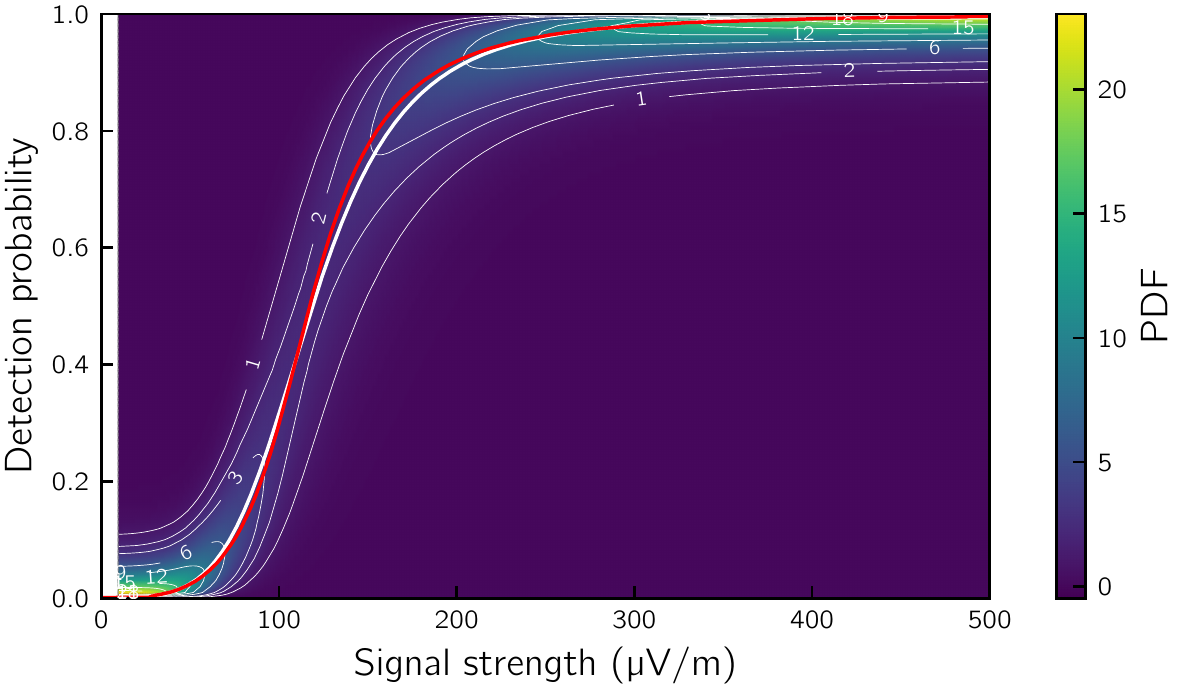}
  \caption{The bivariate spline-interpolation of the convoluted probability density of the signal detection estimated for a grid of signals ranging from $10\,$\textmu V/m to $500\,$\textmu V/m with $10\,$\textmu V/m steps (this function corresponds to Equation~\ref{eq:pdf-convoluted}). 
  The thick white line shows the mode of the convoluted probability density. 
  The red line indicates the position of the mode before performing the convolution (the red line on this plot corresponds to the thick white line in Figure~\ref{fig:pdf}).}
  \label{fig:pdf-convoluted}
\end{figure}

To use the probability density for the signal detection in an individual antenna together with the previously described radio LDF model, we perform a convolution of this probability density for the signal detection with the uncertainty of LDF, which we model with the Gaussian function centered at a given signal strength and standard deviation of $\approx 14.5\,$\% of that strength ($\sigma\approx0.145\mathscr{E}$)
\begin{equation}
    P(\mathscr{E}) = P_0 \int_0^{\infty}
    \frac{1}{\mathrm{B}(\alpha(\mathscr{E}), \beta(\mathscr{E}))}\,
    p^{\alpha(\mathscr{E}) - 1} (1-p)^{\beta(\mathscr{E}) -1}\,
    \exp\left(-\frac{(\xi - \mathscr{E})^2}{2\sigma^2(\mathscr{E})}\right)
    \dd \xi.
    \label{eq:pdf-convoluted}
\end{equation}
The normalization $P_0$ we find numerically. 
To improve performance in further computations, we computed the convoluted densities on a grid of sample points ranging
from $10$ to $500\,$\textmu V/m with $10\,$\textmu V/m steps and interpolated them with a bicubic spline.
Figure~\ref{fig:pdf-convoluted} shows the resulting probability density.
This probability density reflects the signal detection properties of the averaged antenna for signals coming from all directions present in the simulations.

It is worthwhile to note that the probability density changes its meaning after this convolution. 
Before the convolution, a slice for a given abscissa means the probability density to detect a signal with a given strength; after the convolution, the meaning changes to the probability to detect a signal predicted to be of a given strength by the footprint model,  but the actual signal strength could be anywhere within the uncertainties of the prediction.

\subsection{Detection Probability for an Array of Radio Antennas}
With the models described in the previous sections, we can estimate the detection efficiency for a single antenna. 
However, for radio arrays, usually the coincident detection in several antennas is required, and the number of required antennas may depend on the goal of a specific analysis, e.g., three antennas will be sufficient for an approximate reconstruction of the arrival direction, but the reconstruction of $X_{\text{max}}$ will require more antennas with signal, depending on the desired reconstruction precision.
Therefore, this section describes the final step of the model: the probability density to observe a shower with the antenna array, requiring the coincident detection of signals at several antennas.
We treat this component of the efficiency model probabilistically, too, in the same way as the previous components.

We use two different, alternative approaches for this final step of the model, probabilistic calculations and Monte Carlo simulations, and compare them with each other.

\paragraph{Approach with Probabilistic Calculations.}
The basis of the probabilistic calculations for estimation of the detection efficiency is the probabilistic understanding of the air-shower detection process by an array of antennas. 
Appearance or not appearance of a signal at a given antenna is treated as an independent event (``event'' in a probabilistic sense, not as synonym for an observed air-shower). 
The computation of the detection efficiency is based on the calculation of the probabilities of all situations that lead to the detection of the air shower, i.e., those situations with at least the pre-required number of antennas with signal. 

Due to the fact that any antenna can either detect or not detect signal with a certain probability, we consider the probability to observe a given number $n$ of signals from a shower as sum of probabilities to observe all combinations of antennas leading to the observation of $n$ signals in total.
The joint probability of the situation that the first $n$ antennas detect a signal has the following form
\begin{equation}
  p^{(n)} =
  p_1\, p_2\, p_3 \dots p_{n-1}\, p_n\,
  \bar{p}_{n+1}\, \bar{p}_{N-2}\, \bar{p}_{N-1}\, \bar{p}_N.
\end{equation}
The symbols $p_i$ denote the probability densities to detect a signal by $i$-th antenna (this quantity corresponds to the probability density expressed by Equation~\ref{eq:pdf-convoluted} taken at a given signal strength), and $\bar{p}$ denote the probabilities of the non-detection obtained by the complement rule: $\bar{p} = 1-p$.
The total probability to observe $n$ signals over the entire array is the joint probability of all independent events
\begin{equation}
\begin{aligned}
  p^{(n)} = p_1\, p_2\, p_3
            \dots      p_{n-1}\,       p_n\,        \bar{p}_{n+1}
            \dots \bar{p}_{N-n}\, \bar{p}_{N-n+1}\, \bar{p}_{N-n+2}
            \dots \bar{p}_{n-2}\, \bar{p}_{N-1}\,   \bar{p}_{N}& + \\
            p_1\, p_2\, p_3
            \dots      p_{n-1}\,  \bar{p}_n\,            p_{n+1}
            \dots \bar{p}_{N-n}\, \bar{p}_{N-n+1}\, \bar{p}_{N-n+2}
            \dots \bar{p}_{n-2}\, \bar{p}_{N-1}\,   \bar{p}_{N}& + \\
            &\dots \\
            \bar{p}_1\, \bar{p}_2\, \bar{p}_3
            \dots \bar{p}_{n-1}\, \bar{p}_n\,       \bar{p}_{n+1}
            \dots      p_{N-n}\,  \bar{p}_{N-n+1}\,      p_{N-n+2}
            \dots      p_{n-2}\,       p_{N-1}\,         p_{N}& + \\
            \bar{p}_1\, \bar{p}_2\, \bar{p}_3
            \dots \bar{p}_{n-1}\, \bar{p}_n\,      \bar{p}_{n+1}
            \dots \bar{p}_{N-n}\,      p_{N-n+1}\,      p_{N-n+2}
            \dots      p_{n-2}\,       p_{N-1}\,        p_{N}&,
\end{aligned}
\end{equation}
or shortly
\begin{equation}
  p^{(n)} = \sum_{i=1}^{\binom{N}{n}} p_i^{(n)}.
\end{equation}
It is easy to formally write the probability of the detection condition with the introduced notation. 
If we assume that such a detection condition consists in the requirement of at least $m$ antennas with signals, the detection probability is defined by the following equation
\begin{equation}
  P = \sum_{i=1}^{\binom{N}{m}}   p_i^{(m)}   +
      \sum_{i=1}^{\binom{N}{m+1}} p_i^{(m+1)} + \dots +
      \sum_{i=1}^{\binom{N}{N-1}} p_i^{(N1)}  +
      \sum_{i=1}^{\binom{N}{N}}   p_i^{(N)}.
\end{equation}
This equation is correct, however, it is not feasible to use it for practical computation due to the large number of required operations.  
Since usually the number of the required signals in the detection condition is much smaller than total number of antennas of the array, it is more feasible to compute the detection probability via the complement of all situations which do not lead to a detection
\begin{equation}
  P = 1 - \left(
      \sum_{i=1}^{\binom{N}{0}}   p_i^{(0)}   +
      \sum_{i=1}^{\binom{N}{1}}   p_i^{(1)}   + \dots +
      \sum_{i=1}^{\binom{N}{m-1}} p_i^{(m-1)} \right).
\end{equation}

It is important to recall at this point that each of the $p_i$ factors in the equations above is a probability density function, not a simple number, thus, the algebraic operations need to be performed correspondingly~\cite{Springer1979}.

\begin{figure}[t]
\def\w{0.485}
\includegraphics[width=\w\textwidth]{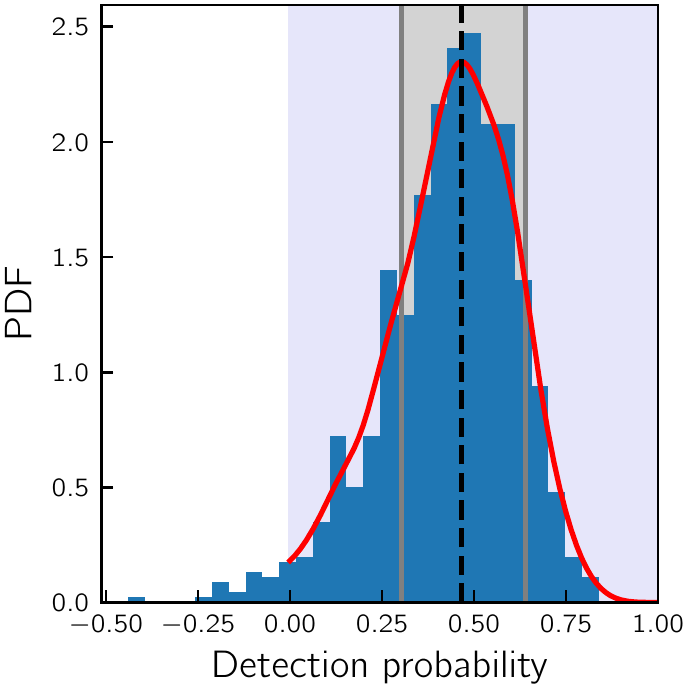}\hfill
\includegraphics[width=\w\textwidth]{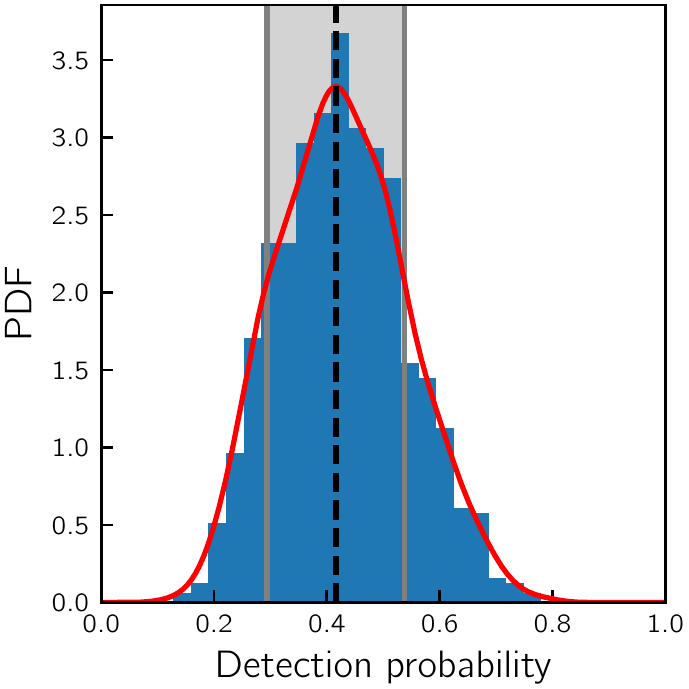}
\caption{Estimation of the probability density function (PDF) for a particular shower taken as example. 
The data are presented in two forms: histogram and kernel density estimator. 
The histogram is normalized to the total number of entries; its binning is obtained with the Freedman Diaconis Estimator. 
The red curve represents the results of the Gaussian kernel density estimation with a bandwidth of $1.06\,\hat{\sigma}\,n^{1/5}$, where $\hat{\sigma}$ is the sample standard deviation. 
The shower core is in the origin of the coordinates, the other shower parameters are $\theta = 30^{\circ}$, $\phi = 270^{\circ}$, $\lg(E/\text{eV})=17.3$, $X_{\text{max}} = 658$\,g/cm$^{2}$.
The vertical dashed black line shows the detection probability, which is the mode of the probability density estimated with the Gaussian kernel; the gray band shows the uncertainty of detection probability and encloses 0.68 of the total region under the probability density.
The detection condition used for this case is at least three antennas with signals among the 63 Tunka-Rex antennas.
  \textit{Left}:~estimation obtained with the probabilistic calculations, $0.466_{-0.162}^{+0.174}$;
  the lavender color marks the physical region of the function domain.
  \textit{Right}:~estimation obtained with the Monte-Carlo approach, $0.416_{-0.122}^{+0.121}$.}
\label{fig:single-event-pdf}
\end{figure}

To practically perform the computations with the probability densities, we use the method of sampling the distributions. 
The idea of the method is as follows. 
We draw a sample from each of the initial distributions for the individual antennas.  
Then, we treat the samples as a certain realization of the probabilities to observe a signal with the antennas. 
A certain realization means that these probabilities become numbers at this point. 
To obtain the detection probability for the array we use the same formulas as shown above, but with the drawn realization of $p_i$ instead.  
By repeating the drawing of samples and conducting the computations with individual realizations, we obtain a sample of the required probability density to detect a shower.  
Then, we use a kernel density estimation with a Gaussian kernel to restore the density itself. 
Figure~\ref{fig:single-event-pdf} (left) shows an example of such a distribution.  
The resulting probability density provides not only the mode of the detection probability, its distribution also provides an estimation of the uncertainty for this mode value.
Usage of the kernel density estimation for reconstruction of the shape of the probability density function mitigates the influence of the number of the samples from which we perform the reconstruction.
For the present version of the model we drew 1000 samples which seems sufficient from the visual investigation of the resulting estimations obtained with both the kernel density estimation and the histogram with the binning obtained with the Freedman-Diaconis rule.

\paragraph{Approach with Monte-Carlo Experiments.}
In some circumstances, such as a relatively large number of antennas required in the detection condition, the calculation method described above performs too slowly. 
To address this problem, we developed an alternative method of Monte Carlo experiments.
It consists of drawing one sample from each of the probability densities to detect a signal in an individual antenna and then run multiple Bernoulli trials with this set of samples. 
The fraction of times when the detection condition is fulfilled provides an estimation of the air-shower detection probability for the particular set of samples.  
By drawing more samples and repeating the procedure we get more estimations of the detection probabilities, and can construct the probability density function for the detection of a given shower. 
The final estimation of the density of the detection probability we obtain again with the kernel density estimation using a Gaussian kernel.
Again, usage of the kernel density estimation mitigates dependence on the sample size.
For the present model we used 1000 samples.
Figure~\ref{fig:single-event-pdf} (right) shows the resulting probability density. 
One can see that both methods provide very close results and could be used interchangeably.

\begin{figure}[t]
  %\hspace{0.05\textwidth}%
  \includegraphics[width=0.485\textwidth]{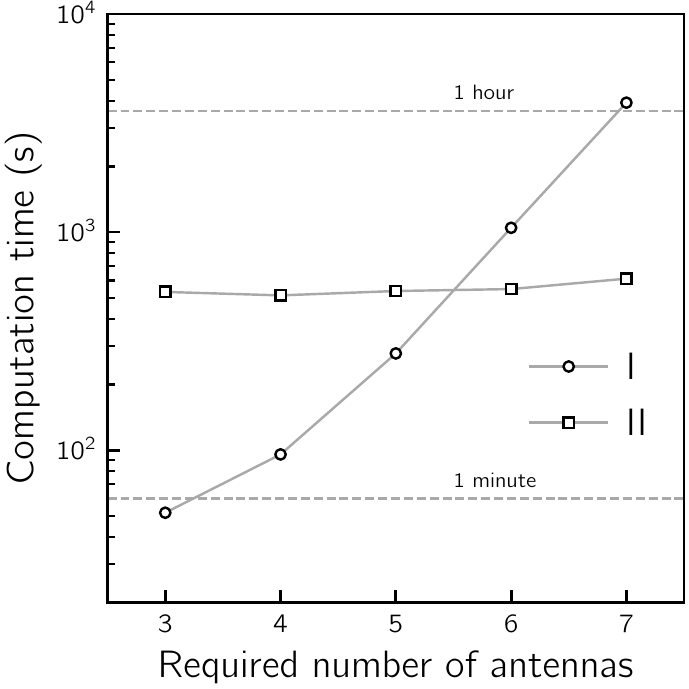}\hfill
    \begin{minipage}[b]{0.455\textwidth}
    \caption{Comparison of the computation time%
    \protect\footnote{CPU: Intel Core i7-4790 @ 3.60\,GHz, memory: 15.6\,GB.}
    for the different number of antennas in the detection condition:\\
    I --- method of probabilistic calculations,
    II --- method of Monte-Carlo experiments.\\
    The test case is computing the averaged efficiency over the Tunka-Rex fiducial area for an event with the following parameters:
    \mbox{$\theta =35^{\circ}$}, 
    \mbox{$\phi =270^{\circ}$},
    \mbox{$E =10^{17.3}$\,eV},
    \mbox{$X_{\text{max}}=658$\,g/cm$^{2}$};
    grid step size of $50\,$m.}%
    \label{fig:time}%
  \end{minipage}
\end{figure}

Since the main motivation for the development of the second method was the large computational complexity of the first one for a large number of antennas in the detection condition, we compared the computing time for both of the methods.  
As a test case we used the computation of the averaged detection efficiency over the fiducial area of the Tunka-Rex antenna array (defined as a circle with a radius of $450\,$m around the center of the array) with multiple showers coming from the same direction and with shower cores distributed on a rectangular grid over the fiducial area of the array.  Figure~\ref{fig:time} shows the resulting computation time for this particular test case for both approaches. 
The benefit of the method of Monte Carlo experiments is clearly the almost constant computation time independent of the number of signals required for the detection condition.

For all results presented further in this work we use the first method of probabilistic calculations applied to a detection condition requiring at least three antennas with signal. 
However, for some efficiency estimations in real case scenarios, the method of Monte Carlo experiments will be highly beneficial, e.g., a high quality $X_\mathrm{max}$ measurement requiring a larger number of antennas.

We can use the methods described above for computation of the detection efficiency for showers initiated by cosmic rays of a
certain energy and with a certain depth of shower maximum.
Figure~\ref{fig:maps} (left) shows an example calculations for the dependence of the detection efficiency on the core position for a given arrival direction, and Figure~\ref{fig:maps} (right) shows a sky map of the efficiency for all arrival directions when averaging over a set of
core positions distributed in a square-grid layout over the fiducial area of Tunka-Rex with step of 50\,m\footnote{For analysis of stability of the results for a range of grid steps see Ref.~\cite{Lenok2022_1000143479}.}.
One can see that the model provides a unique possibility to estimate both the spatial and angular detection efficiencies for a given air shower.
This allows us to select regions of full efficiency for further bias-free analyses of the air-shower measurements.

\section{Aperture of a Radio Array}
\label{sec:aperture}
\begin{figure}[!t]
\def\w{0.485}
\includegraphics[width=\w\textwidth]{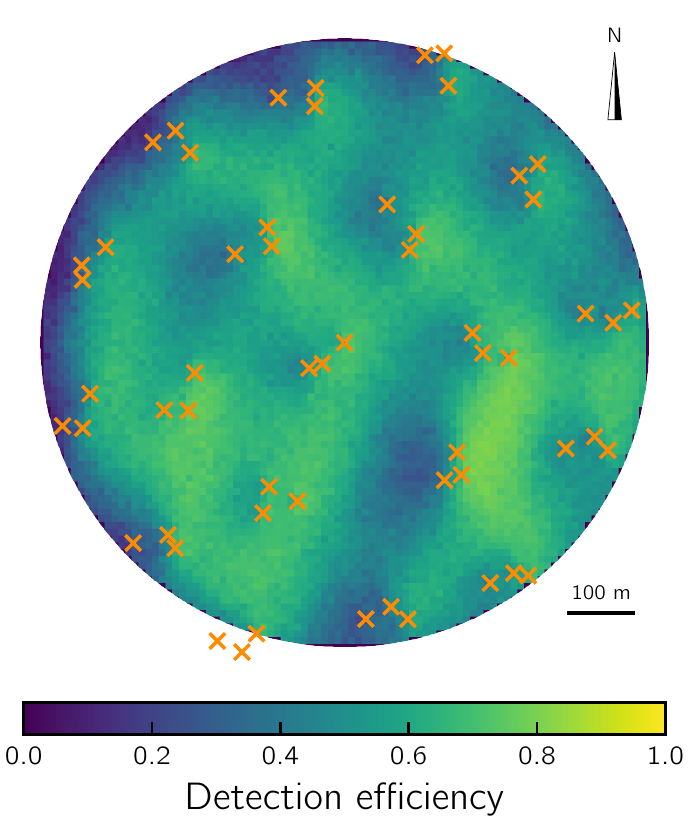}\hfill
\includegraphics[width=\w\textwidth]{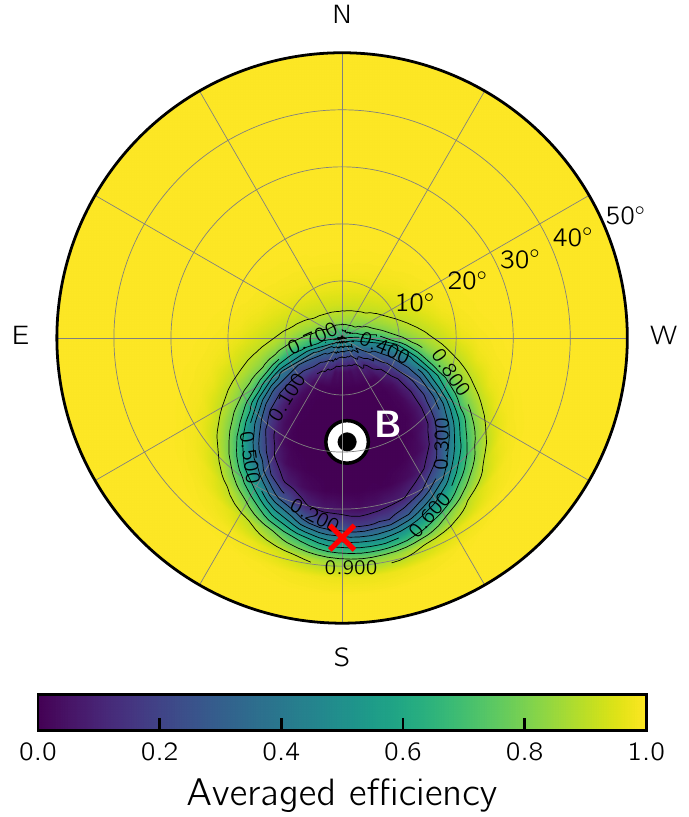}
\caption{The efficiency of Tunka-Rex according to the model developed in this work.
  \textit{Left}: the detection efficiency as function of the core position. 
  The shower has a given energy, $X_{\text{max}}$, and incoming direction (10$^{17.3}$\,eV, 650\,g/cm$^2$, $\theta =35^{\circ}$, $\phi =270^{\circ}$). 
  The arrow in the upper right corner points towards the geographic north. 
  The circular area is the fiducial area of the Tunka-Rex instrument centered at the first antenna position with a fixed radius of $450\,$m; crosses indicate antenna positions.
  \textit{Right}: the detection efficiency averaged over the fiducial area as a function of the incoming direction for $E = 10^{17.3}$\,eV and $X_{\text{max}} = 650$\,g/cm$^2$.
  The black-and-white circle shows the position of the local geomagnetic field. 
  The red cross marks the arrival direction of the left plot ($\theta =35^{\circ}$, $\phi =270^{\circ}$).}
\label{fig:maps}
\end{figure}

The aperture of a cosmic-ray instrument is one of the main characteristics required for reconstruction of the cosmic-ray energy spectrum and mass composition from air-shower measurements.
In contrast to many types of cosmic-ray instruments, radio arrays have a sky region of suppressed efficiency around the direction of the geomagnetic field due to the physics of the emission mechanisms. 
This region can be clearly seen in Figure~\ref{fig:maps} (right).
To avoid biases due to the use of partially efficient sky regions, showers with corresponding arrival directions need to be cut from analyses.

In this section we describe a method to estimate the aperture for the full-efficiency sky region of a radio array. 
To estimate the location of the limited efficiency regions we use the model presented before.

We begin with the formal definition of the aperture and its connection to the detection efficiency and the cosmic-ray flux.
The number of events $N$ in an infinitesimal energy bin ranging from $E$ to $E + \dd E$ observed by a flat cosmic-ray instrument is equal to the cosmic-ray flux $J(E)$ at this energy multiplied by the instrument exposure $\epsilon$.
The latter is an integral of the instrument efficiency $\xi$ integrated over the fiducial area of the instrument $S\rmsub{f}$, the angular sky region selected, $\Omega\rmsub{f}$, and in addition integrated over the operation time $T$
\begin{equation}
  \frac{\dd N(E)}{\dd E} = \epsilon J(E) =
  J(E)
  \int_T \int_{\Omega\rmsub{f}} \int_{S\rmsub{f}}
  \xi \cos\theta\,\dd S\,\dd \Omega\,\dd t.
\end{equation}
The $\cos\theta$ factor here reflects the fact that the considered instrument is flat, which is a good approximation for Tunka-Rex and many other air-shower arrays of similar size.
The efficiency is a function of the cosmic-ray energy $E$, $X_{\text{max}}$, incoming directions $(\theta, \phi)$, and the core position $(x_0, y_0)$: $\xi = \xi(E, X_{\text{max}}, \theta, \phi, x_0, y_0)$.
For simplification of the formulas, we do not list these arguments hereafter.

Under the assumptions that the efficiency of the instrument does not depend on time, at least for selected periods, the integration over time becomes simply a multiplication over the operation time.
The remaining integral holds the name aperture $A$.
\begin{equation}
  A = \int_T \int_{\Omega\rmsub{f}} \int_{S\rmsub{f}}
  \xi \cos\theta\,\dd S\,\dd \Omega\,\dd t =
  T \int_{\Omega\rmsub{f}} \int_{S\rmsub{f}}
  \xi \cos\theta\,\dd S\,\dd \Omega = TA.
\end{equation}
As we can easily estimate the average efficiency over the fiducial area with our model, we transform the aperture integral in the following way to factor out the instrument fiducial area
\begin{equation}
  A = \int_{\Omega\rmsub{f}} \int_{S\rmsub{f}}
  \xi \cos\theta\,\dd S\,\dd \Omega =
  S\rmsub{f} \int_{\Omega\rmsub{f}} \left( \int_{S\rmsub{f}}
  \frac{\xi}{S\rmsub{f}}\,\dd S \right) \cos\theta\,\dd \Omega =
  S\rmsub{f} \int_{\Omega\rmsub{f}}
  \langle\xi\rangle_s
  \cos\theta\,\dd \Omega.
\end{equation}
With this transformation we reduced the initial four-dimensional integral to an integral of only two dimensions of the averaged efficiency over the instrument fiducial area.

The next step is to determine the regions of full efficiency and use them for the integration.

\subsection{Selection of the Full-Efficiency Region}

The efficiency model presented in the previous sections is used to determine the location and size of sky regions with limited efficiency, which are visible in Figure~\ref{fig:maps} (right). 
The threshold of ``full'' efficiency can be defined arbitrarily, but should avoid a significant systematic uncertainty on whatever is the result of a specific analysis (later we will use 98\% as example).
The remaining part of the sky with efficiencies above that threshold is the region of the full efficiency.

As the region with limited efficiency has a close to circular shape, we use a circle with appropriate size and position in the sky to approximate this region in further computations.  
We use the following parametric form of the boarder of this circle
\begin{equation}
  \cos\rho = \cos\theta\cos\theta_0 + \sin\theta\sin\theta_0\cos\phi,
  \label{eq:circle}
\end{equation}
where $\rho$ is the angular radius of the circle and $\theta_0$ is the zenith position of the center of the circle.  
To find these two parameters for air showers with the same properties (i.e., air showers with a given energy, and $X_{\text{max}}$), we compute the efficiency with the model for a pre-defined Gaussian grid on the sphere ($3^{\circ}$ step both in zenith and azimuth) and then interpolate it with linear splines. 
Then we use the obtained linear spline function in a nested minimization procedure to obtain the two parameters of the circle.

\begin{figure}[!t]
\def\w{0.485}
\includegraphics[width=\w\textwidth, valign=c]{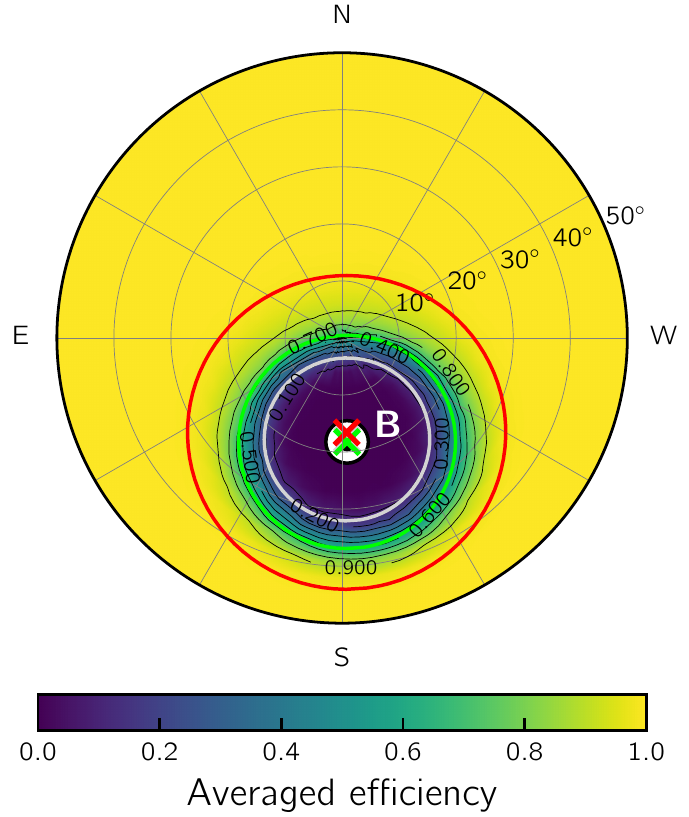}\hfill
\includegraphics[width=\w\textwidth, valign=c]{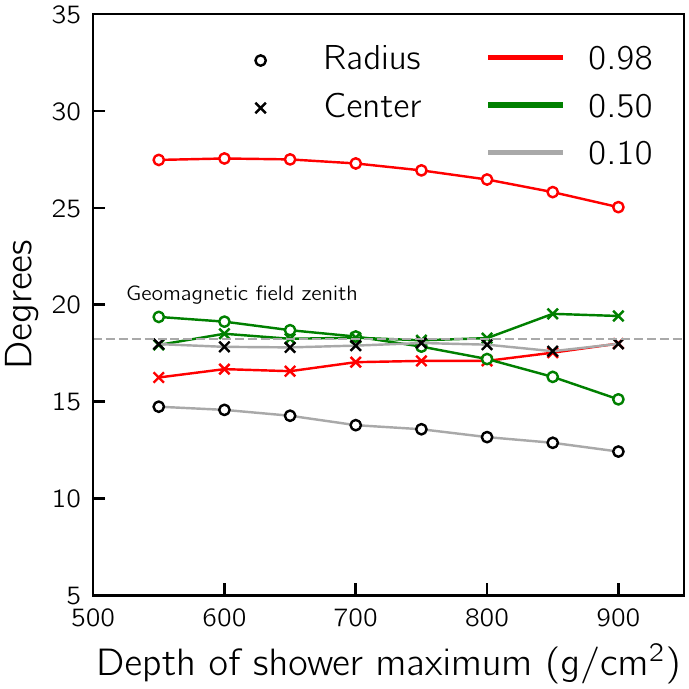}
\caption{Angular behavior of the averaged efficiency for air
  showers with a depth of maximum of 650\,g/cm$^2$ and produced
  by 10$^{17.3}$\,eV cosmic rays.
  \textit{Left}: distribution of the averaged efficiency over
  the sky. The red, green, and gray circles correspond to the
  0.98, 0.5, and 0.1 maximal efficiency regions.
  \textit{Right}: the evolution of the radii and center positions
  of the circles corresponding to the 0.98, 0.50, and 0.1 maximal
  efficiency regions. The size of the 0.98 efficiency circle is
  almost independent of $X_{\text{max}}$.}
  \label{fig:circles}
\end{figure}

We organized the minimization procedure in the following way. 
The external minimization runs over the zenith location of the circle.
The internal minimization looks for the minimal radius of a circle for a given zenith location under the condition that the minimal value of the interpolated efficiency must not be smaller than 98\% on the boarder of the circle. 
Figure~\ref{fig:circles} (left) shows the result of the minimization for a particular shower.

The model presented in the previous section enables us to study both, the spatial and angular dependence, of the detection efficiency as a function of the energy and $X_{\text{max}}$.
The $X_{\text{max}}$ range of interest implicitly contains the information on the mass composition of the cosmic rays, since the model of the radio footprint does not explicitly dependent on the mass of the primary particle.
For purposes of the aperture estimation presented here, we studied how the region of  limited efficiency evolves with changing $X_{\text{max}}$ at a constant energy.
Figure~\ref{fig:circles} (right) briefly summarizes this study.
We found that the size and location of this region changes only marginally over a wide range of $X_{\text{max}}$.
Thus, we conclude that for practical applications a single reference value of $X_{\text{max}}$ can be used to estimate the region of full efficiency, which means that for each energy of interest one sky map is sufficient.

\subsection{Evaluation of the Aperture Integral}

To evaluate the aperture integral, we developed a semi-analytical method. 
The main achievement of the method consists in the conversion of the initial two dimensional aperture integral into a one dimensional one that can be solved numerically with high precision.
On first view, the problem of the aperture calculation from a known full efficiency region might seem simple. 
However, for radio arrays, which have a region of the suppressed efficiency in the sky, this computation requires a two-dimensional numerical integration on a sphere.
This is a complex problem with not many approaches available to date because the numerical integration over a sphere is related to the currently unsolved mathematical problems of a homogeneous distribution of points over a sphere~\cite{Beentjes2016QUADRATUREOA}. 
In this regard, the present method of reducing the two-dimensional problem of the aperture calculation into a one-dimensional one is an important step forward.

We begin with the remaining aperture integral without the fiducial area factor
\begin{equation}
  A_{\Omega} =
  \int_{\Omega\rmsub{f}}
  \langle\xi\rangle_{\text{s}}
  \cos\theta\,\dd \Omega =
  \int_0^{2\pi}\int_0^{\theta\rmsub{max}}
  \langle\xi\rangle_{\text{s}}\cos\theta\sin\theta\,\dd\theta\,\dd\phi.
\end{equation}
This integration should be performed only over the sky region of full
efficiency, which is the entire sky without the approximately circular region of suppressed efficiency. 
By definition, the averaged efficiency in the full efficiency region is one, $\langle\xi\rangle_{\text{s}}=1$, or marginally smaller since it is common to accept efficiency values slightly below one in practical applications. 
For computation of the integral, we split it into two parts. 
From the integral over the full observed sky we subtract the region of suppressed efficiency
\begin{equation}
  A_{\Omega} =
  \int_0^{2\pi} \int_0^{\theta\rmsub{max}}
  \cos\theta\sin\theta\,\dd\theta\,\dd\phi -
  \int_0^{\theta\rmsub{max}} \int_{\phi_1(\theta)}^{\phi_2(\theta)}
  \cos\theta\sin\theta\,\dd\theta\,\dd\phi.
\end{equation}
The solution for the first integral is known and equals 
$\pi\left(1-\cos^2\theta\rmsub{max}\right)$.

To solve the second integral we express the azimuth angle from the
equation of the boarder of the circle (\ref{eq:circle})
\begin{equation}
  \phi = \pm\arccos
  \frac{\cos\rho-\cos\theta\cos\theta_0}
       {\sin\theta\sin\theta_0}
\end{equation}
and place it in the limits of the integral. 
We obtain the following limits
\begin{equation}
  \begin{aligned}
    \phi_1(\theta) &= 0,\\
    \phi_2(\theta) &= \arccos
    \frac{\cos\rho-\cos\theta\cos\theta_0}
         {\sin\theta\sin\theta_0}.
  \end{aligned}
\end{equation}
The plus-minus sign leads to a factor of two in front of the integral due to the symmetry of the efficiency suppressed region.
By applying all these transformations, we reduce the two dimensional integral into a one dimensional integral
\begin{equation}
  \int_0^{\theta\rmsub{max}} \int_{\phi_1(\theta)}^{\phi_2(\theta)}
  \cos\theta\sin\theta\,\dd\theta\,\dd\phi =
  2 \int_0^{\theta\rmsub{max}}\arccos\
  \frac{\cos\rho-\cos\theta\cos\theta_0}
         {\sin\theta\sin\theta_0}
         \cos\theta\sin\theta\,\dd\theta.
\end{equation}

The combination of this result with the known solution for the integral over the entire sky gives the final result for the aperture integral
\begin{equation}
  A_{\Omega} = \pi(1-\cos\theta\rmsub{max}) -
  2 \int_0^{\theta\rmsub{max}}\arccos\
  \frac{\cos\rho-\cos\theta\cos\theta_0}
         {\sin\theta\sin\theta_0}
         \cos\theta\sin\theta\,\dd\theta.
\end{equation}
The remaining one dimensional integral can easily be evaluated numerically.

\section{Validation of the Model}

To check the performance of the efficiency model we validate it against the efficiency estimated from Monte Carlo simulations.%
\footnote{For a validation against measured data see Reference~\cite{Lenok2022_1000143479}.}

\begin{figure}[t]
    \begin{minipage}[b]{0.455\textwidth}
    \caption{Comparison of the detection probability estimated
      with the model against the detection probability determined
      through multiple processing of Monte-Carlo simulations with
      different noise samples.}
    \label{fig:validation}
  \end{minipage}\hfill
    \includegraphics[width=0.485\textwidth]{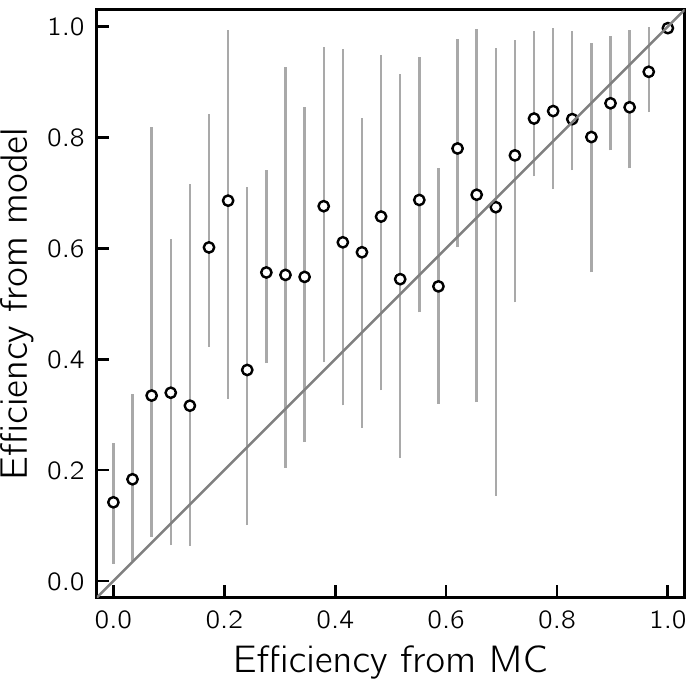}
\end{figure}

The idea behind the estimation of the shower detection efficiency with simulations consists in analyzing the same events multiple times with different measured noise samples added to the radio pulses simulated by CoREAS. 
Then, the fraction of times an event passes the detection condition gives an estimation of the detection efficiency for this event.
For this work we processed each of the events in the simulation set 30 times which is sufficient for obtaining stable results.
From the many simulated events, we formed groups of events with the same Monte Carlo efficiency, and estimated the efficiency with the model for comparison (Figure~\ref{fig:validation}).
The points show the mean values of the model-predicted efficiency for the groups of events with a given Monte Carlo efficiency. 
The error bars indicate the uncertainties of the underlying distributions and represent the range between the 16\% and 84\% percentiles.
The detection condition used for this comparison is at least three antennas with signals.

The comparison shows a very good agreement in the region of high efficiencies which is fully sufficient for reliable detection of the sky regions of full efficiency.
The model may be modified in the future for better performance in the region of intermediate efficiencies.

\section{Conclusion}

We have presented a new model for the estimation of the detection
efficiency of a radio array for cosmic-ray air showers and a method how this model can be used for estimation of the aperture.
The model is built following an explicit probabilistic approach in which we treat each stage of the detection process separately by a corresponding probability density function. 
The final efficiency model is a combination of these probability density functions.

The model addresses challenges arising when estimating the detection efficiency of a radio array with the conventional approach of simulating the array operation by processing many Monte Carlo simulation and applying the detector response. 
One of the main challenges in this approach is the need to generate a sufficient number of air-shower simulations which is difficult in case of radio arrays because of the large computational complexity of the simulation of the radio emission of air showers. 
A simulation-driven estimation of the efficiency as, e.g., done by LOFAR \cite{2015ICRC...34..368B, 2017ICRC...35..499B, 2021PhRvD.103j2006C}, requires the generation of tens of simulated showers for each measured shower.
Although for building the model presented in this work some air-shower simulations were required, too, their number is limited.
Once the model is set, we can study any core positions, incoming directions, energies, and depths of shower maximum without limitations and with very little additional computing time.

The description of the spatial distribution of the air-shower radio emission in the model comes from the LDF used in the reconstruction procedure of the instrument, Tunka-Rex, in our case.
While the model presented here was developed for Tunka-Rex, it has a generic nature and can be applied to any other radio array detecting air showers. 
For doing so, some components of the model should be appropriately modified, namely, the description of the radio footprint and the detection efficiency of the individual antenna. 
Also, in case of using different detection conditions, e.g., topological constrains of the radio footprint in addition to a minimum number of antennas with signal, these need to be incorporated in the model. 

To check the model, we validated it against Monte Carlo simulations which provide the most reliable estimation of the detection efficiency for air showers with given macroparameters.
The comparison revealed that the developed model is in good agreement with the simulations especially for high values of the detection efficiency.  
Some discrepancies can be seen for intermediate efficiency values, which may be due to the simplifications implied in the model, e.g., the radio footprint on ground is only approximated by the LDF used, and even for showers of same $X_{\text{max}}$ the average radio amplitude differs by a few percent depending on the mass of the primary particle \cite{Tunka-Rex:2016nto}. 
Nonetheless, these simplifications do not hamper the application of the model to determine regions of full efficiency in the sky, which are a necessary input for many bias-free analyses in cosmic-ray physics.

\acknowledgments

This work has been supported by the German Academic Exchange Service (DAAD, personal grant No. 91657437). In preparation of this work we used calculations performed on the ForHLR-II cluster. We thank Dmitriy Kostunin, Andreas Haungs, and Tim Huege for useful discussions, and Agnieszka Leszczy\'nska for reading the final manuscript and providing useful comments.

\bibliography{ref-efficiency-paper}{}
\bibliographystyle{JHEP}

\end{document}